\documentstyle [12pt]{article}
\begin{document}

\begin{center}
{\Large {\bf Comparison of meanfield and Monte Carlo approaches to dynamic
hysteresis in Ising ferromagnets}}
\end{center}

\bigskip

\begin{center}
{\large Muktish Acharyya$^*$}
\end{center}
\vspace {0.1 cm}
\begin{center}
{\it Institute for Theoretical Physics, University of Cologne, 50923 Cologne, 
Germany.}
\end{center}

\bigskip
\bigskip
\bigskip

\noindent {\bf Abstract:} The dynamical hysteresis is studied in the 
kinetic Ising model in the presence of a sinusoidal magnetic field
both by Monte Carlo simulation and by solving the dynamical meanfield
equation for the averaged magnetisation. The frequency variations of
the dynamic coercive field are studied below the critical temperature.
In both the cases, it shows a power law frequency variation however it
becomes frequency independent in the low frequency regime for the mean
field case.

\bigskip
\bigskip
\bigskip
\bigskip
\bigskip
\bigskip

\leftline {\bf Keywords: Ising model, Hysteresis, Monte Carlo simulation}
\leftline {\bf PACS Numbers: 05.50 +q}

\bigskip
\bigskip
\bigskip
\bigskip

\leftline {---------------------------------------}
\leftline {\bf $^*$E-mail:muktish@thp.uni-koeln.de}

\newpage

\leftline {\bf I. Introduction}

The hysteresis in ferromagnetic materials is a common interest of theoretical
[1-13]
and experimental [14-16] research. 
How the hysteresis loop area depends, on frequency
and amplitude of the oscillating field and the temperature of the system, is
quite important to know since it has crucial technological importance (e.g.,
magnetic recording media). More than a decade has been spent to study
theoretically [1-13] the intriguing mechanism of hysteresis in the magnetic
model systems for oscillating fields. 
The main outcome of these research is that the loop area ($A = \oint m dh$)
scales as a power law ($A \sim f^a h_0^b$) 
with respect to frequency ($f$) and the amplitude ($h_0$) of the
applied magnetic field. Though there are some disagreement in the values
of the exponents estimated, the qualitative behaviour (power law) 
of the hysteresis loop
area are almost same in various model systems. 

The theoretical studies can be divided into two classes: (i) the mean field
studies [11-13], by solving the 
dynamical meanfield equations where the effects of
fluctuations are not taken into account. In these cases, the hysteresis loop
area scales as a power law with frequency  
apart from a residual loop area in the zero frequency limit. Mathematically,
it can be written as $A = A_0 + C f^a h_0^b$. (ii) The Monte Carlo
and other studies [1-10], where
the effects of fluctuations are taken into account gives the same type of 
power law variation with the absence of any residual loop area 
(i.e., $A_0 = 0$) in the zero
frequency limit. 

There are some experimental efforts [14-16] to study the scaling behaviour of
the hysteresis loop area. The hysteresis has been studied mostly in 
ultrathin ferromagnetic films by magneto-optic Kerr effect. The power law
frequency variation has been observed [14], however the exponent values 
observed [15] are
very small suggesting a possible logarithmic frequency variation. 
Since, the hysteresis loop area linearly scales with the coercive field
(loop area is four times the coercive field times the remanent magnetisation)
the frequency variation of the coercive field has also been studied 
[16] in 
ultrathin (0.8 nm) ferromagnetic films. The coercive field has been found to
be linear in the logarithm of the rate of change of the magnetic field 
(in some sense it is equivalent to the frequency).
The meanfield 
(without any thermal fluctuation)  equations are used to analyse those
experimental data. The theoretical prediction agrees well with the
experimental observation. However, in the zero frequency limit, the loop area
does not vanish.

Comprehending the theoretical facts, it is observed that the meanfield type
models gives nonzero loop area (or coercive field) in the zero frequency limit
however the models which allows the effects of thermal fluctuations gives the
hysteresis loop area which vanishes in the zero frequency 
limit following a power
law variation.  There is no possible explanation why the meanfield results
are giving a nonzero residual hysteresis loop area in the zero frequency
limit. In this paper, to get the possible explanation, the dynamic coercivity
has been studied as a function of frequency in the kinetic Ising model both
by Monte Carlo simulation (to take into account the effects of fluctuation)
and by solving the dynamical meanfield equation (fluctuations are absent).

The paper is organised as follows: in section II the model and the numerical
schemes are discussed, the results are given in section III, in section IV
the results are analysed by simple arguments.

\bigskip

\leftline {\bf II. The model and simulation}
\leftline {\it (a) Monte Carlo simulation}

The total energy (at time $t$) of a nearest neighbour ferromagnetic Ising
model with homogeneous and unit interaction energy can be written as
\begin{equation}
H = -(1/2) \sum_i h_i(t)\sigma_i(t), ~~h_i(t) = \sum_j \sigma_j(t) + h(t)
\end{equation}

\noindent where $\sigma_i(t) = \pm1$ and $j$ runs over the nearest neighbour
of site $i$. The local field (at site $i$) $h_i(t)$ has an external field
part $h(t)$, which is oscillating sinusoidally in time
\begin{equation}
h(t) = h_0 \sin(2\pi f t),
\end{equation}

\noindent where $h_0$ and $f$ are the amplitude and frequency of the 
oscillating field.

According to heat-bath dynamics, the orientation probability
$p_i(t)$ for the spin $\sigma_i(t)$ at time $t$ to show against
the direction of the field is given as
\begin{equation}
p_i(t) = {{e^{-h_i(t)/K_BT}} \over {e^{h_i(t)/K_BT} + e^{-h_i(t)/K_BT}}},
\end{equation}

\noindent where $K_B$ is the Boltzmann constant which has been taken
equal to unity for simplicity. The spin $\sigma_i(t)$ is oriented as

\begin{equation}
\sigma_i(t+1) = {\rm Sign} [r_i(t) - p_i(t)]
\end{equation}

\noindent where $r_i(t)$ are independent random fractions drawn from the
uniform distribution between 0 and 1.

In the simulation, a square lattice ($L\times L$) is considered under
periodic boundary conditions. The initial condition is all spins are
up (i.e., $\sigma_i(t=0) = 1$, for all $i$). The multispin coding technique
is employed here to store 10 spins in a computer
word consisting of 32 bits. 10 spins are updated simultaneously
(or parallel) by a single command. All words are 
updated sequentially and one full scan over the entire lattice consists
of one Monte Carlo step (MCS) per spin. This is the unit of time in the
simulation. The instantaneous magnetisation ($m(t) = (1/L^2) \sum_i \sigma_i
(t)$) is calculated easily. The dynamic hysteresis loop (or $m-h$ loop)
is nothing but the plot of the magnetisation ($m(t)$) and the
magnetic field ($h(t)$). One such loop is displayed in Fig. 1. The
coercive field is the minimum field required for the sign reversal of the
magnetisation. In Fig.1, A is the field necessary (in the negative
direction) for the sign reversal
of the magnetisation. Similarly, B is the field required to push the system
from negative magnetisation state to the 
positive magnetisation state. This 
field is the coercive field ($h_c$). Theoretically, A and B
should be equal in magnitude. However, in practice, due to the presence of
fluctuations they can differ by a small amount. Here the coercive field is
measured by taking the arithmetic mean of these two values (A and B). 
Initially, in the transient region of time, some hysteresis loops are thrown 
out
until a stable and symmetric loop has been obtained. For a fixed frequency
($f$) the coercive field is calculated by averaging over 10 different 
random samples. This simulation is performed in a SUN workstation cluster
and the computational speed recorded is 7.14 Million updates per second.

\bigskip

\leftline {\it (b) Solution of dynamical meanfield equation}

The meanfield dynamical equation of Ising ferromagnet in the presence of
time varying magnetic field is [12]
\begin{equation}
{dm \over dt} = -m + {\rm tanh} \left( {{m(t) + h(t)} \over K_BT} \right),
\end{equation}

\noindent where the external time varying field $h(t)$ has the above form
(2). To study the dynamic hysteresis, this equation has been solved for
$m(t)$ by fourth order Runge-Kutta method by taking the
initial condition  $m(t=0) = 1.0$. In this case also, few transient loops
were discarded to have stable and symmetric loop. From the stable and 
symmetric hysteresis loop the coercive field is measured as the value of
the magnetic field when the magnetisation changes its sign.

\bigskip

\leftline {\bf III. Results}
\leftline {\it (a) Monte Carlo results}

In the Monte Carlo simulation, a square lattice of linear size $L = 1000$ has
been considered. The temperature has been set at 20 percent below the
critical temperature ($T_c$) and kept fixed throughout the study. The
coercive fields are measured for different frequencies ranging from 
5$\times 10^{-5}$ to $10^{-2}$. Frequency is measured in the units of
inverse of MCS. For example, for a frequency $f = 10^{-2}$, 100 
MCS are required to have a closed loop (complete cycle of the oscillating
magnetic field). One such loop (for $f = 0.01$ and $T = 0.8T_c$) 
is shown in Fig.1. The frequency ($f$) variation of dynamic coercive field
 ($h_c$) for a fixed temperature ($T = 0.8T_c$) is plotted in a double
logarithmically in Fig. 2. The variation is clearly a power law over
a resonably wide range of frequency and $h_c(f \to 0) = 0$. The exponent
estimated is $0.33\pm0.02$.

\bigskip
\leftline {\it (b) Meanfield results}

In the meanfield study, the temperature is also kept at 20 percent below
$T_c$ and the frequency ranges from $10^{-5}$ to 1. the frequency variation
of the dynamic coercive field ($h_c$) is shown in a double logarithmic plot
in Fig. 2. Unlike the earlier case (Monte Carlo results), the frequency
variation is a power law type in the high frequency range, however it
becomes frequency independent over a quite wide range in the low frequency
regime and $h_c(f \to 0) \neq 0$. The exponent, of the power law (in the
high frequency regime), estimated here is 0.45.

\bigskip
\leftline {\bf IV. Concluding remarks}

In conclusion, the dynamical hysteresis is studied, in the kinetic Ising
model below the critical temperature and in the presence of a sinusoidal 
(in time) magnetic field, both by Monte Carlo simulation (using multispin
coding and heat bath algorithm) and by solving the meanfield dynamical
equation for the averaged magnetisation. In both the cases, the numerical
results of the frequency variations of the dynamic coercive field, show
a power law frequency variation. In the meanfield case the coercive field
becomes frequency independent in the low frequency regime.

The dissimilarity (MC and MF results) in the qualitative behaviour of
the frequency variations of the dynamic coercivity can be explained
from the Landau type double well free energy. The coercive field is
the field needed to bring the system from one well to another. In the
case of Monte Carlo study, where the thermal fluctuations are present,
there is always a nonzero finite probability that the system can go to
the other well being driven by thermal fluctuation. In this case, even 
a vanishingly small field can help the system to cross the free
energy barrier via the {\it nucleation} process if one waits for a sufficiently
long time. In this respect, the 
coercivity can be thought as a limit of metastability. This
situation means: in the zero frequency limit, the coercive field
vanishes. On the other hand, in the case of meanfield study, the thermal
fluctuation is absent. As a result, below $T_c$, unless a finite amount
(non zero) of field is applied, the system will not jump from one well to
the other, even if one waits for an infinitely long time. This is equivalent
to say that below $T_c$, in the static limit ($f \to 0$), the coercive
field is not zero. The numerical results are indeed qualitatively consistent
with this argument. Thermal fluctuations are playing the major role to 
make this kind of distinction in the
frequency variations of the coercive field in
the static ($f \to 0$) limit.

The theoretical observations [11,13] indirectly support the results and the
argument by showing nonzero residual loop area in the zero frequency limit.
The experimental observations [16] are also in agreement with this argument.
The logarithmic variation [16,17] 
is quite slow variation, essentially independent of
frequency and the coercive field does not vanish as frequency tend to zero.

\bigskip
\leftline {\bf Acknowledgments:} {\bf SFB 341} is gratefully acknowledged for
financial support.

\newpage
%%%%%%%%%%END OF THE TEXT%%%%%%%%%%%%%%%%%%%%%%%%%%%%%%%%%%%%%%%%%%%

%%%%%%%%%%%%THREE FIGURES%%%%%%%%%%%%%%%%%%%%%%%%%%%%%%%%%%%%%%%%%%%%
%FIGURE-1
% GNUPLOT: LaTeX picture
\setlength{\unitlength}{0.240900pt}
\ifx\plotpoint\undefined\newsavebox{\plotpoint}\fi
\sbox{\plotpoint}{\rule[-0.200pt]{0.400pt}{0.400pt}}%
\begin{picture}(1049,900)(0,0)
\font\gnuplot=cmr10 at 10pt
\gnuplot
\sbox{\plotpoint}{\rule[-0.200pt]{0.400pt}{0.400pt}}%
\put(220.0,495.0){\rule[-0.200pt]{184.288pt}{0.400pt}}
\put(603.0,113.0){\rule[-0.200pt]{0.400pt}{184.048pt}}
\put(220.0,177.0){\rule[-0.200pt]{4.818pt}{0.400pt}}
\put(198,177){\makebox(0,0)[r]{-1}}
\put(965.0,177.0){\rule[-0.200pt]{4.818pt}{0.400pt}}
\put(220.0,336.0){\rule[-0.200pt]{4.818pt}{0.400pt}}
\put(198,336){\makebox(0,0)[r]{-0.5}}
\put(965.0,336.0){\rule[-0.200pt]{4.818pt}{0.400pt}}
\put(220.0,495.0){\rule[-0.200pt]{4.818pt}{0.400pt}}
\put(198,495){\makebox(0,0)[r]{0}}
\put(965.0,495.0){\rule[-0.200pt]{4.818pt}{0.400pt}}
\put(220.0,654.0){\rule[-0.200pt]{4.818pt}{0.400pt}}
\put(198,654){\makebox(0,0)[r]{0.5}}
\put(965.0,654.0){\rule[-0.200pt]{4.818pt}{0.400pt}}
\put(220.0,813.0){\rule[-0.200pt]{4.818pt}{0.400pt}}
\put(198,813){\makebox(0,0)[r]{1}}
\put(965.0,813.0){\rule[-0.200pt]{4.818pt}{0.400pt}}
\put(220.0,113.0){\rule[-0.200pt]{0.400pt}{4.818pt}}
\put(220,68){\makebox(0,0){-6}}
\put(220.0,857.0){\rule[-0.200pt]{0.400pt}{4.818pt}}
\put(348.0,113.0){\rule[-0.200pt]{0.400pt}{4.818pt}}
\put(348,68){\makebox(0,0){-4}}
\put(348.0,857.0){\rule[-0.200pt]{0.400pt}{4.818pt}}
\put(475.0,113.0){\rule[-0.200pt]{0.400pt}{4.818pt}}
\put(475,68){\makebox(0,0){-2}}
\put(475.0,857.0){\rule[-0.200pt]{0.400pt}{4.818pt}}
\put(603.0,113.0){\rule[-0.200pt]{0.400pt}{4.818pt}}
\put(603,68){\makebox(0,0){0}}
\put(603.0,857.0){\rule[-0.200pt]{0.400pt}{4.818pt}}
\put(730.0,113.0){\rule[-0.200pt]{0.400pt}{4.818pt}}
\put(730,68){\makebox(0,0){2}}
\put(730.0,857.0){\rule[-0.200pt]{0.400pt}{4.818pt}}
\put(858.0,113.0){\rule[-0.200pt]{0.400pt}{4.818pt}}
\put(858,68){\makebox(0,0){4}}
\put(858.0,857.0){\rule[-0.200pt]{0.400pt}{4.818pt}}
\put(985.0,113.0){\rule[-0.200pt]{0.400pt}{4.818pt}}
\put(985,68){\makebox(0,0){6}}
\put(985.0,857.0){\rule[-0.200pt]{0.400pt}{4.818pt}}
\put(220.0,113.0){\rule[-0.200pt]{184.288pt}{0.400pt}}
\put(985.0,113.0){\rule[-0.200pt]{0.400pt}{184.048pt}}
\put(220.0,877.0){\rule[-0.200pt]{184.288pt}{0.400pt}}
\put(45,495){\makebox(0,0){$m(t)$}}
\put(602,23){\makebox(0,0){$h(t)$}}
\put(615,521){\makebox(0,0)[l]{O}}
\put(480,521){\makebox(0,0)[l]{A}}
\put(714,521){\makebox(0,0)[l]{B}}
\put(252,813){\makebox(0,0)[l]{$L=1000$}}
\put(252,750){\makebox(0,0)[l]{$T=0.8T_c$}}
\put(252,686){\makebox(0,0)[l]{$f=0.01$}}
\put(220.0,113.0){\rule[-0.200pt]{0.400pt}{184.048pt}}
\put(921,813){\usebox{\plotpoint}}
\put(720,811.67){\rule{4.336pt}{0.400pt}}
\multiput(729.00,812.17)(-9.000,-1.000){2}{\rule{2.168pt}{0.400pt}}
\put(738.0,813.0){\rule[-0.200pt]{44.085pt}{0.400pt}}
\put(682,810.67){\rule{4.577pt}{0.400pt}}
\multiput(691.50,811.17)(-9.500,-1.000){2}{\rule{2.289pt}{0.400pt}}
\put(662,809.67){\rule{4.818pt}{0.400pt}}
\multiput(672.00,810.17)(-10.000,-1.000){2}{\rule{2.409pt}{0.400pt}}
\put(642,808.67){\rule{4.818pt}{0.400pt}}
\multiput(652.00,809.17)(-10.000,-1.000){2}{\rule{2.409pt}{0.400pt}}
\multiput(631.07,807.95)(-4.034,-0.447){3}{\rule{2.633pt}{0.108pt}}
\multiput(636.53,808.17)(-13.534,-3.000){2}{\rule{1.317pt}{0.400pt}}
\multiput(613.87,804.94)(-2.967,-0.468){5}{\rule{2.200pt}{0.113pt}}
\multiput(618.43,805.17)(-16.434,-4.000){2}{\rule{1.100pt}{0.400pt}}
\multiput(597.43,800.93)(-1.286,-0.488){13}{\rule{1.100pt}{0.117pt}}
\multiput(599.72,801.17)(-17.717,-8.000){2}{\rule{0.550pt}{0.400pt}}
\multiput(579.83,792.92)(-0.526,-0.495){33}{\rule{0.522pt}{0.119pt}}
\multiput(580.92,793.17)(-17.916,-18.000){2}{\rule{0.261pt}{0.400pt}}
\multiput(561.92,771.68)(-0.496,-1.186){37}{\rule{0.119pt}{1.040pt}}
\multiput(562.17,773.84)(-20.000,-44.841){2}{\rule{0.400pt}{0.520pt}}
\multiput(541.92,717.96)(-0.496,-3.249){37}{\rule{0.119pt}{2.660pt}}
\multiput(542.17,723.48)(-20.000,-122.479){2}{\rule{0.400pt}{1.330pt}}
\multiput(521.92,580.48)(-0.495,-6.161){35}{\rule{0.119pt}{4.942pt}}
\multiput(522.17,590.74)(-19.000,-219.742){2}{\rule{0.400pt}{2.471pt}}
\multiput(502.92,356.60)(-0.495,-4.282){35}{\rule{0.119pt}{3.468pt}}
\multiput(503.17,363.80)(-19.000,-152.801){2}{\rule{0.400pt}{1.734pt}}
\multiput(483.92,207.63)(-0.495,-0.895){33}{\rule{0.119pt}{0.811pt}}
\multiput(484.17,209.32)(-18.000,-30.316){2}{\rule{0.400pt}{0.406pt}}
\put(449,177.17){\rule{3.700pt}{0.400pt}}
\multiput(459.32,178.17)(-10.320,-2.000){2}{\rule{1.850pt}{0.400pt}}
\put(701.0,812.0){\rule[-0.200pt]{4.577pt}{0.400pt}}
\put(284.0,177.0){\rule[-0.200pt]{39.748pt}{0.400pt}}
\put(467,176.67){\rule{4.336pt}{0.400pt}}
\multiput(467.00,176.17)(9.000,1.000){2}{\rule{2.168pt}{0.400pt}}
\put(284.0,177.0){\rule[-0.200pt]{44.085pt}{0.400pt}}
\put(504,177.67){\rule{4.577pt}{0.400pt}}
\multiput(504.00,177.17)(9.500,1.000){2}{\rule{2.289pt}{0.400pt}}
\put(523,178.67){\rule{4.818pt}{0.400pt}}
\multiput(523.00,178.17)(10.000,1.000){2}{\rule{2.409pt}{0.400pt}}
\put(543,179.67){\rule{4.818pt}{0.400pt}}
\multiput(543.00,179.17)(10.000,1.000){2}{\rule{2.409pt}{0.400pt}}
\multiput(563.00,181.61)(4.034,0.447){3}{\rule{2.633pt}{0.108pt}}
\multiput(563.00,180.17)(13.534,3.000){2}{\rule{1.317pt}{0.400pt}}
\multiput(582.00,184.60)(2.967,0.468){5}{\rule{2.200pt}{0.113pt}}
\multiput(582.00,183.17)(16.434,4.000){2}{\rule{1.100pt}{0.400pt}}
\multiput(603.00,188.59)(1.286,0.488){13}{\rule{1.100pt}{0.117pt}}
\multiput(603.00,187.17)(17.717,8.000){2}{\rule{0.550pt}{0.400pt}}
\multiput(623.00,196.58)(0.526,0.495){33}{\rule{0.522pt}{0.119pt}}
\multiput(623.00,195.17)(17.916,18.000){2}{\rule{0.261pt}{0.400pt}}
\multiput(642.58,214.00)(0.496,1.186){37}{\rule{0.119pt}{1.040pt}}
\multiput(641.17,214.00)(20.000,44.841){2}{\rule{0.400pt}{0.520pt}}
\multiput(662.58,261.00)(0.496,3.223){37}{\rule{0.119pt}{2.640pt}}
\multiput(661.17,261.00)(20.000,121.521){2}{\rule{0.400pt}{1.320pt}}
\multiput(682.58,388.00)(0.495,6.188){35}{\rule{0.119pt}{4.963pt}}
\multiput(681.17,388.00)(19.000,220.699){2}{\rule{0.400pt}{2.482pt}}
\multiput(701.58,619.00)(0.495,4.282){35}{\rule{0.119pt}{3.468pt}}
\multiput(700.17,619.00)(19.000,152.801){2}{\rule{0.400pt}{1.734pt}}
\multiput(720.58,779.00)(0.495,0.895){33}{\rule{0.119pt}{0.811pt}}
\multiput(719.17,779.00)(18.000,30.316){2}{\rule{0.400pt}{0.406pt}}
\put(738,811.17){\rule{3.700pt}{0.400pt}}
\multiput(738.00,810.17)(10.320,2.000){2}{\rule{1.850pt}{0.400pt}}
\put(485.0,178.0){\rule[-0.200pt]{4.577pt}{0.400pt}}
\put(756.0,813.0){\rule[-0.200pt]{39.748pt}{0.400pt}}
\put(220,495){\usebox{\plotpoint}}
\put(220.0,495.0){\rule[-0.200pt]{184.288pt}{0.400pt}}
\put(603,113){\usebox{\plotpoint}}
\put(603.0,113.0){\rule[-0.200pt]{0.400pt}{184.048pt}}
\end{picture}

\bigskip

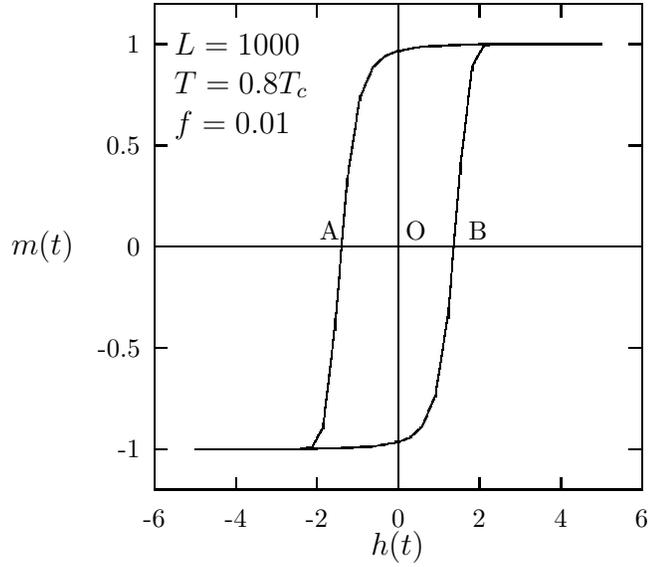
\noindent {\bf Fig.1.} A typical hysteresis loop obtained from the MC
simulation.

\bigskip

%FIGURE-2
% GNUPLOT: LaTeX picture
\setlength{\unitlength}{0.240900pt}
\ifx\plotpoint\undefined\newsavebox{\plotpoint}\fi
\sbox{\plotpoint}{\rule[-0.200pt]{0.400pt}{0.400pt}}%
\begin{picture}(1049,900)(0,0)
\font\gnuplot=cmr10 at 10pt
\gnuplot
\sbox{\plotpoint}{\rule[-0.200pt]{0.400pt}{0.400pt}}%
\put(220.0,113.0){\rule[-0.200pt]{4.818pt}{0.400pt}}
\put(198,113){\makebox(0,0)[r]{0.01}}
\put(965.0,113.0){\rule[-0.200pt]{4.818pt}{0.400pt}}
\put(220.0,213.0){\rule[-0.200pt]{2.409pt}{0.400pt}}
\put(975.0,213.0){\rule[-0.200pt]{2.409pt}{0.400pt}}
\put(220.0,271.0){\rule[-0.200pt]{2.409pt}{0.400pt}}
\put(975.0,271.0){\rule[-0.200pt]{2.409pt}{0.400pt}}
\put(220.0,313.0){\rule[-0.200pt]{2.409pt}{0.400pt}}
\put(975.0,313.0){\rule[-0.200pt]{2.409pt}{0.400pt}}
\put(220.0,345.0){\rule[-0.200pt]{2.409pt}{0.400pt}}
\put(975.0,345.0){\rule[-0.200pt]{2.409pt}{0.400pt}}
\put(220.0,371.0){\rule[-0.200pt]{2.409pt}{0.400pt}}
\put(975.0,371.0){\rule[-0.200pt]{2.409pt}{0.400pt}}
\put(220.0,394.0){\rule[-0.200pt]{2.409pt}{0.400pt}}
\put(975.0,394.0){\rule[-0.200pt]{2.409pt}{0.400pt}}
\put(220.0,413.0){\rule[-0.200pt]{2.409pt}{0.400pt}}
\put(975.0,413.0){\rule[-0.200pt]{2.409pt}{0.400pt}}
\put(220.0,430.0){\rule[-0.200pt]{2.409pt}{0.400pt}}
\put(975.0,430.0){\rule[-0.200pt]{2.409pt}{0.400pt}}
\put(220.0,445.0){\rule[-0.200pt]{4.818pt}{0.400pt}}
\put(198,445){\makebox(0,0)[r]{0.1}}
\put(965.0,445.0){\rule[-0.200pt]{4.818pt}{0.400pt}}
\put(220.0,545.0){\rule[-0.200pt]{2.409pt}{0.400pt}}
\put(975.0,545.0){\rule[-0.200pt]{2.409pt}{0.400pt}}
\put(220.0,603.0){\rule[-0.200pt]{2.409pt}{0.400pt}}
\put(975.0,603.0){\rule[-0.200pt]{2.409pt}{0.400pt}}
\put(220.0,645.0){\rule[-0.200pt]{2.409pt}{0.400pt}}
\put(975.0,645.0){\rule[-0.200pt]{2.409pt}{0.400pt}}
\put(220.0,677.0){\rule[-0.200pt]{2.409pt}{0.400pt}}
\put(975.0,677.0){\rule[-0.200pt]{2.409pt}{0.400pt}}
\put(220.0,703.0){\rule[-0.200pt]{2.409pt}{0.400pt}}
\put(975.0,703.0){\rule[-0.200pt]{2.409pt}{0.400pt}}
\put(220.0,726.0){\rule[-0.200pt]{2.409pt}{0.400pt}}
\put(975.0,726.0){\rule[-0.200pt]{2.409pt}{0.400pt}}
\put(220.0,745.0){\rule[-0.200pt]{2.409pt}{0.400pt}}
\put(975.0,745.0){\rule[-0.200pt]{2.409pt}{0.400pt}}
\put(220.0,762.0){\rule[-0.200pt]{2.409pt}{0.400pt}}
\put(975.0,762.0){\rule[-0.200pt]{2.409pt}{0.400pt}}
\put(220.0,777.0){\rule[-0.200pt]{4.818pt}{0.400pt}}
\put(198,777){\makebox(0,0)[r]{1}}
\put(965.0,777.0){\rule[-0.200pt]{4.818pt}{0.400pt}}
\put(220.0,877.0){\rule[-0.200pt]{2.409pt}{0.400pt}}
\put(975.0,877.0){\rule[-0.200pt]{2.409pt}{0.400pt}}
\put(220.0,113.0){\rule[-0.200pt]{0.400pt}{4.818pt}}
\put(220,68){\makebox(0,0){1e-05}}
\put(220.0,857.0){\rule[-0.200pt]{0.400pt}{4.818pt}}
\put(266.0,113.0){\rule[-0.200pt]{0.400pt}{2.409pt}}
\put(266.0,867.0){\rule[-0.200pt]{0.400pt}{2.409pt}}
\put(327.0,113.0){\rule[-0.200pt]{0.400pt}{2.409pt}}
\put(327.0,867.0){\rule[-0.200pt]{0.400pt}{2.409pt}}
\put(358.0,113.0){\rule[-0.200pt]{0.400pt}{2.409pt}}
\put(358.0,867.0){\rule[-0.200pt]{0.400pt}{2.409pt}}
\put(373.0,113.0){\rule[-0.200pt]{0.400pt}{4.818pt}}
\put(373,68){\makebox(0,0){0.0001}}
\put(373.0,857.0){\rule[-0.200pt]{0.400pt}{4.818pt}}
\put(419.0,113.0){\rule[-0.200pt]{0.400pt}{2.409pt}}
\put(419.0,867.0){\rule[-0.200pt]{0.400pt}{2.409pt}}
\put(480.0,113.0){\rule[-0.200pt]{0.400pt}{2.409pt}}
\put(480.0,867.0){\rule[-0.200pt]{0.400pt}{2.409pt}}
\put(511.0,113.0){\rule[-0.200pt]{0.400pt}{2.409pt}}
\put(511.0,867.0){\rule[-0.200pt]{0.400pt}{2.409pt}}
\put(526.0,113.0){\rule[-0.200pt]{0.400pt}{4.818pt}}
\put(526,68){\makebox(0,0){0.001}}
\put(526.0,857.0){\rule[-0.200pt]{0.400pt}{4.818pt}}
\put(572.0,113.0){\rule[-0.200pt]{0.400pt}{2.409pt}}
\put(572.0,867.0){\rule[-0.200pt]{0.400pt}{2.409pt}}
\put(633.0,113.0){\rule[-0.200pt]{0.400pt}{2.409pt}}
\put(633.0,867.0){\rule[-0.200pt]{0.400pt}{2.409pt}}
\put(664.0,113.0){\rule[-0.200pt]{0.400pt}{2.409pt}}
\put(664.0,867.0){\rule[-0.200pt]{0.400pt}{2.409pt}}
\put(679.0,113.0){\rule[-0.200pt]{0.400pt}{4.818pt}}
\put(679,68){\makebox(0,0){0.01}}
\put(679.0,857.0){\rule[-0.200pt]{0.400pt}{4.818pt}}
\put(725.0,113.0){\rule[-0.200pt]{0.400pt}{2.409pt}}
\put(725.0,867.0){\rule[-0.200pt]{0.400pt}{2.409pt}}
\put(786.0,113.0){\rule[-0.200pt]{0.400pt}{2.409pt}}
\put(786.0,867.0){\rule[-0.200pt]{0.400pt}{2.409pt}}
\put(817.0,113.0){\rule[-0.200pt]{0.400pt}{2.409pt}}
\put(817.0,867.0){\rule[-0.200pt]{0.400pt}{2.409pt}}
\put(832.0,113.0){\rule[-0.200pt]{0.400pt}{4.818pt}}
\put(832,68){\makebox(0,0){0.1}}
\put(832.0,857.0){\rule[-0.200pt]{0.400pt}{4.818pt}}
\put(878.0,113.0){\rule[-0.200pt]{0.400pt}{2.409pt}}
\put(878.0,867.0){\rule[-0.200pt]{0.400pt}{2.409pt}}
\put(939.0,113.0){\rule[-0.200pt]{0.400pt}{2.409pt}}
\put(939.0,867.0){\rule[-0.200pt]{0.400pt}{2.409pt}}
\put(970.0,113.0){\rule[-0.200pt]{0.400pt}{2.409pt}}
\put(970.0,867.0){\rule[-0.200pt]{0.400pt}{2.409pt}}
\put(985.0,113.0){\rule[-0.200pt]{0.400pt}{4.818pt}}
\put(985,68){\makebox(0,0){1}}
\put(985.0,857.0){\rule[-0.200pt]{0.400pt}{4.818pt}}
\put(220.0,113.0){\rule[-0.200pt]{184.288pt}{0.400pt}}
\put(985.0,113.0){\rule[-0.200pt]{0.400pt}{184.048pt}}
\put(220.0,877.0){\rule[-0.200pt]{184.288pt}{0.400pt}}
\put(45,495){\makebox(0,0){$h_c$}}
\put(602,23){\makebox(0,0){$f$}}
\put(220.0,113.0){\rule[-0.200pt]{0.400pt}{184.048pt}}
\put(985,740){\raisebox{-.8pt}{\makebox(0,0){$\bullet$}}}
\put(939,700){\raisebox{-.8pt}{\makebox(0,0){$\bullet$}}}
\put(893,649){\raisebox{-.8pt}{\makebox(0,0){$\bullet$}}}
\put(832,590){\raisebox{-.8pt}{\makebox(0,0){$\bullet$}}}
\put(786,549){\raisebox{-.8pt}{\makebox(0,0){$\bullet$}}}
\put(740,511){\raisebox{-.8pt}{\makebox(0,0){$\bullet$}}}
\put(679,469){\raisebox{-.8pt}{\makebox(0,0){$\bullet$}}}
\put(633,444){\raisebox{-.8pt}{\makebox(0,0){$\bullet$}}}
\put(587,424){\raisebox{-.8pt}{\makebox(0,0){$\bullet$}}}
\put(526,405){\raisebox{-.8pt}{\makebox(0,0){$\bullet$}}}
\put(480,396){\raisebox{-.8pt}{\makebox(0,0){$\bullet$}}}
\put(434,389){\raisebox{-.8pt}{\makebox(0,0){$\bullet$}}}
\put(373,384){\raisebox{-.8pt}{\makebox(0,0){$\bullet$}}}
\put(327,381){\raisebox{-.8pt}{\makebox(0,0){$\bullet$}}}
\put(281,380){\raisebox{-.8pt}{\makebox(0,0){$\bullet$}}}
\put(220,378){\raisebox{-.8pt}{\makebox(0,0){$\bullet$}}}
\multiput(347.00,113.60)(0.481,0.468){5}{\rule{0.500pt}{0.113pt}}
\multiput(347.00,112.17)(2.962,4.000){2}{\rule{0.250pt}{0.400pt}}
\multiput(351.00,117.59)(0.569,0.485){11}{\rule{0.557pt}{0.117pt}}
\multiput(351.00,116.17)(6.844,7.000){2}{\rule{0.279pt}{0.400pt}}
\multiput(359.00,124.59)(0.494,0.488){13}{\rule{0.500pt}{0.117pt}}
\multiput(359.00,123.17)(6.962,8.000){2}{\rule{0.250pt}{0.400pt}}
\multiput(367.00,132.59)(0.569,0.485){11}{\rule{0.557pt}{0.117pt}}
\multiput(367.00,131.17)(6.844,7.000){2}{\rule{0.279pt}{0.400pt}}
\multiput(375.59,139.00)(0.485,0.569){11}{\rule{0.117pt}{0.557pt}}
\multiput(374.17,139.00)(7.000,6.844){2}{\rule{0.400pt}{0.279pt}}
\multiput(382.00,147.59)(0.494,0.488){13}{\rule{0.500pt}{0.117pt}}
\multiput(382.00,146.17)(6.962,8.000){2}{\rule{0.250pt}{0.400pt}}
\multiput(390.00,155.59)(0.569,0.485){11}{\rule{0.557pt}{0.117pt}}
\multiput(390.00,154.17)(6.844,7.000){2}{\rule{0.279pt}{0.400pt}}
\multiput(398.59,162.00)(0.485,0.569){11}{\rule{0.117pt}{0.557pt}}
\multiput(397.17,162.00)(7.000,6.844){2}{\rule{0.400pt}{0.279pt}}
\multiput(405.00,170.59)(0.569,0.485){11}{\rule{0.557pt}{0.117pt}}
\multiput(405.00,169.17)(6.844,7.000){2}{\rule{0.279pt}{0.400pt}}
\multiput(413.00,177.59)(0.494,0.488){13}{\rule{0.500pt}{0.117pt}}
\multiput(413.00,176.17)(6.962,8.000){2}{\rule{0.250pt}{0.400pt}}
\multiput(421.00,185.59)(0.569,0.485){11}{\rule{0.557pt}{0.117pt}}
\multiput(421.00,184.17)(6.844,7.000){2}{\rule{0.279pt}{0.400pt}}
\multiput(429.59,192.00)(0.485,0.569){11}{\rule{0.117pt}{0.557pt}}
\multiput(428.17,192.00)(7.000,6.844){2}{\rule{0.400pt}{0.279pt}}
\multiput(436.00,200.59)(0.569,0.485){11}{\rule{0.557pt}{0.117pt}}
\multiput(436.00,199.17)(6.844,7.000){2}{\rule{0.279pt}{0.400pt}}
\multiput(444.00,207.59)(0.494,0.488){13}{\rule{0.500pt}{0.117pt}}
\multiput(444.00,206.17)(6.962,8.000){2}{\rule{0.250pt}{0.400pt}}
\multiput(452.00,215.59)(0.569,0.485){11}{\rule{0.557pt}{0.117pt}}
\multiput(452.00,214.17)(6.844,7.000){2}{\rule{0.279pt}{0.400pt}}
\multiput(460.59,222.00)(0.485,0.569){11}{\rule{0.117pt}{0.557pt}}
\multiput(459.17,222.00)(7.000,6.844){2}{\rule{0.400pt}{0.279pt}}
\multiput(467.00,230.59)(0.494,0.488){13}{\rule{0.500pt}{0.117pt}}
\multiput(467.00,229.17)(6.962,8.000){2}{\rule{0.250pt}{0.400pt}}
\multiput(475.00,238.59)(0.569,0.485){11}{\rule{0.557pt}{0.117pt}}
\multiput(475.00,237.17)(6.844,7.000){2}{\rule{0.279pt}{0.400pt}}
\multiput(483.59,245.00)(0.485,0.569){11}{\rule{0.117pt}{0.557pt}}
\multiput(482.17,245.00)(7.000,6.844){2}{\rule{0.400pt}{0.279pt}}
\multiput(490.00,253.59)(0.569,0.485){11}{\rule{0.557pt}{0.117pt}}
\multiput(490.00,252.17)(6.844,7.000){2}{\rule{0.279pt}{0.400pt}}
\multiput(498.00,260.59)(0.494,0.488){13}{\rule{0.500pt}{0.117pt}}
\multiput(498.00,259.17)(6.962,8.000){2}{\rule{0.250pt}{0.400pt}}
\multiput(506.00,268.59)(0.569,0.485){11}{\rule{0.557pt}{0.117pt}}
\multiput(506.00,267.17)(6.844,7.000){2}{\rule{0.279pt}{0.400pt}}
\multiput(514.59,275.00)(0.485,0.569){11}{\rule{0.117pt}{0.557pt}}
\multiput(513.17,275.00)(7.000,6.844){2}{\rule{0.400pt}{0.279pt}}
\multiput(521.00,283.59)(0.569,0.485){11}{\rule{0.557pt}{0.117pt}}
\multiput(521.00,282.17)(6.844,7.000){2}{\rule{0.279pt}{0.400pt}}
\multiput(529.00,290.59)(0.494,0.488){13}{\rule{0.500pt}{0.117pt}}
\multiput(529.00,289.17)(6.962,8.000){2}{\rule{0.250pt}{0.400pt}}
\multiput(537.00,298.59)(0.569,0.485){11}{\rule{0.557pt}{0.117pt}}
\multiput(537.00,297.17)(6.844,7.000){2}{\rule{0.279pt}{0.400pt}}
\multiput(545.59,305.00)(0.485,0.569){11}{\rule{0.117pt}{0.557pt}}
\multiput(544.17,305.00)(7.000,6.844){2}{\rule{0.400pt}{0.279pt}}
\multiput(552.00,313.59)(0.494,0.488){13}{\rule{0.500pt}{0.117pt}}
\multiput(552.00,312.17)(6.962,8.000){2}{\rule{0.250pt}{0.400pt}}
\multiput(560.00,321.59)(0.569,0.485){11}{\rule{0.557pt}{0.117pt}}
\multiput(560.00,320.17)(6.844,7.000){2}{\rule{0.279pt}{0.400pt}}
\multiput(568.59,328.00)(0.485,0.569){11}{\rule{0.117pt}{0.557pt}}
\multiput(567.17,328.00)(7.000,6.844){2}{\rule{0.400pt}{0.279pt}}
\multiput(575.00,336.59)(0.569,0.485){11}{\rule{0.557pt}{0.117pt}}
\multiput(575.00,335.17)(6.844,7.000){2}{\rule{0.279pt}{0.400pt}}
\multiput(583.00,343.59)(0.494,0.488){13}{\rule{0.500pt}{0.117pt}}
\multiput(583.00,342.17)(6.962,8.000){2}{\rule{0.250pt}{0.400pt}}
\multiput(591.00,351.59)(0.569,0.485){11}{\rule{0.557pt}{0.117pt}}
\multiput(591.00,350.17)(6.844,7.000){2}{\rule{0.279pt}{0.400pt}}
\multiput(599.59,358.00)(0.485,0.569){11}{\rule{0.117pt}{0.557pt}}
\multiput(598.17,358.00)(7.000,6.844){2}{\rule{0.400pt}{0.279pt}}
\multiput(606.00,366.59)(0.569,0.485){11}{\rule{0.557pt}{0.117pt}}
\multiput(606.00,365.17)(6.844,7.000){2}{\rule{0.279pt}{0.400pt}}
\multiput(614.00,373.59)(0.494,0.488){13}{\rule{0.500pt}{0.117pt}}
\multiput(614.00,372.17)(6.962,8.000){2}{\rule{0.250pt}{0.400pt}}
\multiput(622.00,381.59)(0.569,0.485){11}{\rule{0.557pt}{0.117pt}}
\multiput(622.00,380.17)(6.844,7.000){2}{\rule{0.279pt}{0.400pt}}
\multiput(630.59,388.00)(0.485,0.569){11}{\rule{0.117pt}{0.557pt}}
\multiput(629.17,388.00)(7.000,6.844){2}{\rule{0.400pt}{0.279pt}}
\multiput(637.00,396.59)(0.494,0.488){13}{\rule{0.500pt}{0.117pt}}
\multiput(637.00,395.17)(6.962,8.000){2}{\rule{0.250pt}{0.400pt}}
\multiput(645.00,404.59)(0.569,0.485){11}{\rule{0.557pt}{0.117pt}}
\multiput(645.00,403.17)(6.844,7.000){2}{\rule{0.279pt}{0.400pt}}
\multiput(653.59,411.00)(0.485,0.569){11}{\rule{0.117pt}{0.557pt}}
\multiput(652.17,411.00)(7.000,6.844){2}{\rule{0.400pt}{0.279pt}}
\multiput(660.00,419.59)(0.569,0.485){11}{\rule{0.557pt}{0.117pt}}
\multiput(660.00,418.17)(6.844,7.000){2}{\rule{0.279pt}{0.400pt}}
\multiput(668.00,426.59)(0.494,0.488){13}{\rule{0.500pt}{0.117pt}}
\multiput(668.00,425.17)(6.962,8.000){2}{\rule{0.250pt}{0.400pt}}
\multiput(676.00,434.59)(0.569,0.485){11}{\rule{0.557pt}{0.117pt}}
\multiput(676.00,433.17)(6.844,7.000){2}{\rule{0.279pt}{0.400pt}}
\multiput(684.59,441.00)(0.485,0.569){11}{\rule{0.117pt}{0.557pt}}
\multiput(683.17,441.00)(7.000,6.844){2}{\rule{0.400pt}{0.279pt}}
\multiput(691.00,449.59)(0.569,0.485){11}{\rule{0.557pt}{0.117pt}}
\multiput(691.00,448.17)(6.844,7.000){2}{\rule{0.279pt}{0.400pt}}
\multiput(699.00,456.59)(0.494,0.488){13}{\rule{0.500pt}{0.117pt}}
\multiput(699.00,455.17)(6.962,8.000){2}{\rule{0.250pt}{0.400pt}}
\multiput(707.00,464.59)(0.569,0.485){11}{\rule{0.557pt}{0.117pt}}
\multiput(707.00,463.17)(6.844,7.000){2}{\rule{0.279pt}{0.400pt}}
\multiput(715.59,471.00)(0.485,0.569){11}{\rule{0.117pt}{0.557pt}}
\multiput(714.17,471.00)(7.000,6.844){2}{\rule{0.400pt}{0.279pt}}
\multiput(722.00,479.59)(0.494,0.488){13}{\rule{0.500pt}{0.117pt}}
\multiput(722.00,478.17)(6.962,8.000){2}{\rule{0.250pt}{0.400pt}}
\multiput(730.00,487.59)(0.569,0.485){11}{\rule{0.557pt}{0.117pt}}
\multiput(730.00,486.17)(6.844,7.000){2}{\rule{0.279pt}{0.400pt}}
\multiput(738.59,494.00)(0.485,0.569){11}{\rule{0.117pt}{0.557pt}}
\multiput(737.17,494.00)(7.000,6.844){2}{\rule{0.400pt}{0.279pt}}
\multiput(745.00,502.59)(0.569,0.485){11}{\rule{0.557pt}{0.117pt}}
\multiput(745.00,501.17)(6.844,7.000){2}{\rule{0.279pt}{0.400pt}}
\multiput(753.00,509.59)(0.494,0.488){13}{\rule{0.500pt}{0.117pt}}
\multiput(753.00,508.17)(6.962,8.000){2}{\rule{0.250pt}{0.400pt}}
\multiput(761.00,517.59)(0.569,0.485){11}{\rule{0.557pt}{0.117pt}}
\multiput(761.00,516.17)(6.844,7.000){2}{\rule{0.279pt}{0.400pt}}
\multiput(769.59,524.00)(0.485,0.569){11}{\rule{0.117pt}{0.557pt}}
\multiput(768.17,524.00)(7.000,6.844){2}{\rule{0.400pt}{0.279pt}}
\multiput(776.00,532.59)(0.569,0.485){11}{\rule{0.557pt}{0.117pt}}
\multiput(776.00,531.17)(6.844,7.000){2}{\rule{0.279pt}{0.400pt}}
\multiput(784.00,539.59)(0.494,0.488){13}{\rule{0.500pt}{0.117pt}}
\multiput(784.00,538.17)(6.962,8.000){2}{\rule{0.250pt}{0.400pt}}
\multiput(792.00,547.59)(0.569,0.485){11}{\rule{0.557pt}{0.117pt}}
\multiput(792.00,546.17)(6.844,7.000){2}{\rule{0.279pt}{0.400pt}}
\multiput(800.59,554.00)(0.485,0.569){11}{\rule{0.117pt}{0.557pt}}
\multiput(799.17,554.00)(7.000,6.844){2}{\rule{0.400pt}{0.279pt}}
\multiput(807.00,562.59)(0.494,0.488){13}{\rule{0.500pt}{0.117pt}}
\multiput(807.00,561.17)(6.962,8.000){2}{\rule{0.250pt}{0.400pt}}
\multiput(815.00,570.59)(0.569,0.485){11}{\rule{0.557pt}{0.117pt}}
\multiput(815.00,569.17)(6.844,7.000){2}{\rule{0.279pt}{0.400pt}}
\multiput(823.59,577.00)(0.485,0.569){11}{\rule{0.117pt}{0.557pt}}
\multiput(822.17,577.00)(7.000,6.844){2}{\rule{0.400pt}{0.279pt}}
\multiput(830.00,585.59)(0.569,0.485){11}{\rule{0.557pt}{0.117pt}}
\multiput(830.00,584.17)(6.844,7.000){2}{\rule{0.279pt}{0.400pt}}
\multiput(838.00,592.59)(0.494,0.488){13}{\rule{0.500pt}{0.117pt}}
\multiput(838.00,591.17)(6.962,8.000){2}{\rule{0.250pt}{0.400pt}}
\multiput(846.00,600.59)(0.569,0.485){11}{\rule{0.557pt}{0.117pt}}
\multiput(846.00,599.17)(6.844,7.000){2}{\rule{0.279pt}{0.400pt}}
\multiput(854.59,607.00)(0.485,0.569){11}{\rule{0.117pt}{0.557pt}}
\multiput(853.17,607.00)(7.000,6.844){2}{\rule{0.400pt}{0.279pt}}
\multiput(861.00,615.59)(0.569,0.485){11}{\rule{0.557pt}{0.117pt}}
\multiput(861.00,614.17)(6.844,7.000){2}{\rule{0.279pt}{0.400pt}}
\multiput(869.00,622.59)(0.494,0.488){13}{\rule{0.500pt}{0.117pt}}
\multiput(869.00,621.17)(6.962,8.000){2}{\rule{0.250pt}{0.400pt}}
\multiput(877.00,630.59)(0.569,0.485){11}{\rule{0.557pt}{0.117pt}}
\multiput(877.00,629.17)(6.844,7.000){2}{\rule{0.279pt}{0.400pt}}
\multiput(885.59,637.00)(0.485,0.569){11}{\rule{0.117pt}{0.557pt}}
\multiput(884.17,637.00)(7.000,6.844){2}{\rule{0.400pt}{0.279pt}}
\multiput(892.00,645.59)(0.494,0.488){13}{\rule{0.500pt}{0.117pt}}
\multiput(892.00,644.17)(6.962,8.000){2}{\rule{0.250pt}{0.400pt}}
\multiput(900.00,653.59)(0.569,0.485){11}{\rule{0.557pt}{0.117pt}}
\multiput(900.00,652.17)(6.844,7.000){2}{\rule{0.279pt}{0.400pt}}
\multiput(908.59,660.00)(0.485,0.569){11}{\rule{0.117pt}{0.557pt}}
\multiput(907.17,660.00)(7.000,6.844){2}{\rule{0.400pt}{0.279pt}}
\multiput(915.00,668.59)(0.569,0.485){11}{\rule{0.557pt}{0.117pt}}
\multiput(915.00,667.17)(6.844,7.000){2}{\rule{0.279pt}{0.400pt}}
\multiput(923.00,675.59)(0.494,0.488){13}{\rule{0.500pt}{0.117pt}}
\multiput(923.00,674.17)(6.962,8.000){2}{\rule{0.250pt}{0.400pt}}
\multiput(931.00,683.59)(0.569,0.485){11}{\rule{0.557pt}{0.117pt}}
\multiput(931.00,682.17)(6.844,7.000){2}{\rule{0.279pt}{0.400pt}}
\multiput(939.59,690.00)(0.485,0.569){11}{\rule{0.117pt}{0.557pt}}
\multiput(938.17,690.00)(7.000,6.844){2}{\rule{0.400pt}{0.279pt}}
\multiput(946.00,698.59)(0.569,0.485){11}{\rule{0.557pt}{0.117pt}}
\multiput(946.00,697.17)(6.844,7.000){2}{\rule{0.279pt}{0.400pt}}
\multiput(954.00,705.59)(0.494,0.488){13}{\rule{0.500pt}{0.117pt}}
\multiput(954.00,704.17)(6.962,8.000){2}{\rule{0.250pt}{0.400pt}}
\multiput(962.00,713.59)(0.569,0.485){11}{\rule{0.557pt}{0.117pt}}
\multiput(962.00,712.17)(6.844,7.000){2}{\rule{0.279pt}{0.400pt}}
\multiput(970.59,720.00)(0.485,0.569){11}{\rule{0.117pt}{0.557pt}}
\multiput(969.17,720.00)(7.000,6.844){2}{\rule{0.400pt}{0.279pt}}
\multiput(977.00,728.59)(0.494,0.488){13}{\rule{0.500pt}{0.117pt}}
\multiput(977.00,727.17)(6.962,8.000){2}{\rule{0.250pt}{0.400pt}}
\sbox{\plotpoint}{\rule[-0.400pt]{0.800pt}{0.800pt}}%
\put(679,840){\makebox(0,0){$\Diamond$}}
\put(633,808){\makebox(0,0){$\Diamond$}}
\put(587,768){\makebox(0,0){$\Diamond$}}
\put(526,723){\makebox(0,0){$\Diamond$}}
\put(480,690){\makebox(0,0){$\Diamond$}}
\put(434,658){\makebox(0,0){$\Diamond$}}
\put(373,618){\makebox(0,0){$\Diamond$}}
\put(327,590){\makebox(0,0){$\Diamond$}}
\put(250,775){\makebox(0,0)[l]{$T=0.8T_c$}}
\sbox{\plotpoint}{\rule[-0.200pt]{0.400pt}{0.400pt}}%
\put(220,508){\usebox{\plotpoint}}
\multiput(220.00,508.59)(0.821,0.477){7}{\rule{0.740pt}{0.115pt}}
\multiput(220.00,507.17)(6.464,5.000){2}{\rule{0.370pt}{0.400pt}}
\multiput(228.00,513.59)(0.581,0.482){9}{\rule{0.567pt}{0.116pt}}
\multiput(228.00,512.17)(5.824,6.000){2}{\rule{0.283pt}{0.400pt}}
\multiput(235.00,519.59)(0.821,0.477){7}{\rule{0.740pt}{0.115pt}}
\multiput(235.00,518.17)(6.464,5.000){2}{\rule{0.370pt}{0.400pt}}
\multiput(243.00,524.59)(0.671,0.482){9}{\rule{0.633pt}{0.116pt}}
\multiput(243.00,523.17)(6.685,6.000){2}{\rule{0.317pt}{0.400pt}}
\multiput(251.00,530.59)(0.821,0.477){7}{\rule{0.740pt}{0.115pt}}
\multiput(251.00,529.17)(6.464,5.000){2}{\rule{0.370pt}{0.400pt}}
\multiput(259.00,535.59)(0.581,0.482){9}{\rule{0.567pt}{0.116pt}}
\multiput(259.00,534.17)(5.824,6.000){2}{\rule{0.283pt}{0.400pt}}
\multiput(266.00,541.59)(0.821,0.477){7}{\rule{0.740pt}{0.115pt}}
\multiput(266.00,540.17)(6.464,5.000){2}{\rule{0.370pt}{0.400pt}}
\multiput(274.00,546.59)(0.671,0.482){9}{\rule{0.633pt}{0.116pt}}
\multiput(274.00,545.17)(6.685,6.000){2}{\rule{0.317pt}{0.400pt}}
\multiput(282.00,552.59)(0.671,0.482){9}{\rule{0.633pt}{0.116pt}}
\multiput(282.00,551.17)(6.685,6.000){2}{\rule{0.317pt}{0.400pt}}
\multiput(290.00,558.59)(0.710,0.477){7}{\rule{0.660pt}{0.115pt}}
\multiput(290.00,557.17)(5.630,5.000){2}{\rule{0.330pt}{0.400pt}}
\multiput(297.00,563.59)(0.671,0.482){9}{\rule{0.633pt}{0.116pt}}
\multiput(297.00,562.17)(6.685,6.000){2}{\rule{0.317pt}{0.400pt}}
\multiput(305.00,569.59)(0.821,0.477){7}{\rule{0.740pt}{0.115pt}}
\multiput(305.00,568.17)(6.464,5.000){2}{\rule{0.370pt}{0.400pt}}
\multiput(313.00,574.59)(0.581,0.482){9}{\rule{0.567pt}{0.116pt}}
\multiput(313.00,573.17)(5.824,6.000){2}{\rule{0.283pt}{0.400pt}}
\multiput(320.00,580.59)(0.821,0.477){7}{\rule{0.740pt}{0.115pt}}
\multiput(320.00,579.17)(6.464,5.000){2}{\rule{0.370pt}{0.400pt}}
\multiput(328.00,585.59)(0.671,0.482){9}{\rule{0.633pt}{0.116pt}}
\multiput(328.00,584.17)(6.685,6.000){2}{\rule{0.317pt}{0.400pt}}
\multiput(336.00,591.59)(0.821,0.477){7}{\rule{0.740pt}{0.115pt}}
\multiput(336.00,590.17)(6.464,5.000){2}{\rule{0.370pt}{0.400pt}}
\multiput(344.00,596.59)(0.581,0.482){9}{\rule{0.567pt}{0.116pt}}
\multiput(344.00,595.17)(5.824,6.000){2}{\rule{0.283pt}{0.400pt}}
\multiput(351.00,602.59)(0.821,0.477){7}{\rule{0.740pt}{0.115pt}}
\multiput(351.00,601.17)(6.464,5.000){2}{\rule{0.370pt}{0.400pt}}
\multiput(359.00,607.59)(0.671,0.482){9}{\rule{0.633pt}{0.116pt}}
\multiput(359.00,606.17)(6.685,6.000){2}{\rule{0.317pt}{0.400pt}}
\multiput(367.00,613.59)(0.821,0.477){7}{\rule{0.740pt}{0.115pt}}
\multiput(367.00,612.17)(6.464,5.000){2}{\rule{0.370pt}{0.400pt}}
\multiput(375.00,618.59)(0.581,0.482){9}{\rule{0.567pt}{0.116pt}}
\multiput(375.00,617.17)(5.824,6.000){2}{\rule{0.283pt}{0.400pt}}
\multiput(382.00,624.59)(0.821,0.477){7}{\rule{0.740pt}{0.115pt}}
\multiput(382.00,623.17)(6.464,5.000){2}{\rule{0.370pt}{0.400pt}}
\multiput(390.00,629.59)(0.671,0.482){9}{\rule{0.633pt}{0.116pt}}
\multiput(390.00,628.17)(6.685,6.000){2}{\rule{0.317pt}{0.400pt}}
\multiput(398.00,635.59)(0.581,0.482){9}{\rule{0.567pt}{0.116pt}}
\multiput(398.00,634.17)(5.824,6.000){2}{\rule{0.283pt}{0.400pt}}
\multiput(405.00,641.59)(0.821,0.477){7}{\rule{0.740pt}{0.115pt}}
\multiput(405.00,640.17)(6.464,5.000){2}{\rule{0.370pt}{0.400pt}}
\multiput(413.00,646.59)(0.671,0.482){9}{\rule{0.633pt}{0.116pt}}
\multiput(413.00,645.17)(6.685,6.000){2}{\rule{0.317pt}{0.400pt}}
\multiput(421.00,652.59)(0.821,0.477){7}{\rule{0.740pt}{0.115pt}}
\multiput(421.00,651.17)(6.464,5.000){2}{\rule{0.370pt}{0.400pt}}
\multiput(429.00,657.59)(0.581,0.482){9}{\rule{0.567pt}{0.116pt}}
\multiput(429.00,656.17)(5.824,6.000){2}{\rule{0.283pt}{0.400pt}}
\multiput(436.00,663.59)(0.821,0.477){7}{\rule{0.740pt}{0.115pt}}
\multiput(436.00,662.17)(6.464,5.000){2}{\rule{0.370pt}{0.400pt}}
\multiput(444.00,668.59)(0.671,0.482){9}{\rule{0.633pt}{0.116pt}}
\multiput(444.00,667.17)(6.685,6.000){2}{\rule{0.317pt}{0.400pt}}
\multiput(452.00,674.59)(0.821,0.477){7}{\rule{0.740pt}{0.115pt}}
\multiput(452.00,673.17)(6.464,5.000){2}{\rule{0.370pt}{0.400pt}}
\multiput(460.00,679.59)(0.581,0.482){9}{\rule{0.567pt}{0.116pt}}
\multiput(460.00,678.17)(5.824,6.000){2}{\rule{0.283pt}{0.400pt}}
\multiput(467.00,685.59)(0.821,0.477){7}{\rule{0.740pt}{0.115pt}}
\multiput(467.00,684.17)(6.464,5.000){2}{\rule{0.370pt}{0.400pt}}
\multiput(475.00,690.59)(0.671,0.482){9}{\rule{0.633pt}{0.116pt}}
\multiput(475.00,689.17)(6.685,6.000){2}{\rule{0.317pt}{0.400pt}}
\multiput(483.00,696.59)(0.710,0.477){7}{\rule{0.660pt}{0.115pt}}
\multiput(483.00,695.17)(5.630,5.000){2}{\rule{0.330pt}{0.400pt}}
\multiput(490.00,701.59)(0.671,0.482){9}{\rule{0.633pt}{0.116pt}}
\multiput(490.00,700.17)(6.685,6.000){2}{\rule{0.317pt}{0.400pt}}
\multiput(498.00,707.59)(0.821,0.477){7}{\rule{0.740pt}{0.115pt}}
\multiput(498.00,706.17)(6.464,5.000){2}{\rule{0.370pt}{0.400pt}}
\multiput(506.00,712.59)(0.671,0.482){9}{\rule{0.633pt}{0.116pt}}
\multiput(506.00,711.17)(6.685,6.000){2}{\rule{0.317pt}{0.400pt}}
\multiput(514.00,718.59)(0.581,0.482){9}{\rule{0.567pt}{0.116pt}}
\multiput(514.00,717.17)(5.824,6.000){2}{\rule{0.283pt}{0.400pt}}
\multiput(521.00,724.59)(0.821,0.477){7}{\rule{0.740pt}{0.115pt}}
\multiput(521.00,723.17)(6.464,5.000){2}{\rule{0.370pt}{0.400pt}}
\multiput(529.00,729.59)(0.671,0.482){9}{\rule{0.633pt}{0.116pt}}
\multiput(529.00,728.17)(6.685,6.000){2}{\rule{0.317pt}{0.400pt}}
\multiput(537.00,735.59)(0.821,0.477){7}{\rule{0.740pt}{0.115pt}}
\multiput(537.00,734.17)(6.464,5.000){2}{\rule{0.370pt}{0.400pt}}
\multiput(545.00,740.59)(0.581,0.482){9}{\rule{0.567pt}{0.116pt}}
\multiput(545.00,739.17)(5.824,6.000){2}{\rule{0.283pt}{0.400pt}}
\multiput(552.00,746.59)(0.821,0.477){7}{\rule{0.740pt}{0.115pt}}
\multiput(552.00,745.17)(6.464,5.000){2}{\rule{0.370pt}{0.400pt}}
\multiput(560.00,751.59)(0.671,0.482){9}{\rule{0.633pt}{0.116pt}}
\multiput(560.00,750.17)(6.685,6.000){2}{\rule{0.317pt}{0.400pt}}
\multiput(568.00,757.59)(0.710,0.477){7}{\rule{0.660pt}{0.115pt}}
\multiput(568.00,756.17)(5.630,5.000){2}{\rule{0.330pt}{0.400pt}}
\multiput(575.00,762.59)(0.671,0.482){9}{\rule{0.633pt}{0.116pt}}
\multiput(575.00,761.17)(6.685,6.000){2}{\rule{0.317pt}{0.400pt}}
\multiput(583.00,768.59)(0.821,0.477){7}{\rule{0.740pt}{0.115pt}}
\multiput(583.00,767.17)(6.464,5.000){2}{\rule{0.370pt}{0.400pt}}
\multiput(591.00,773.59)(0.671,0.482){9}{\rule{0.633pt}{0.116pt}}
\multiput(591.00,772.17)(6.685,6.000){2}{\rule{0.317pt}{0.400pt}}
\multiput(599.00,779.59)(0.710,0.477){7}{\rule{0.660pt}{0.115pt}}
\multiput(599.00,778.17)(5.630,5.000){2}{\rule{0.330pt}{0.400pt}}
\multiput(606.00,784.59)(0.671,0.482){9}{\rule{0.633pt}{0.116pt}}
\multiput(606.00,783.17)(6.685,6.000){2}{\rule{0.317pt}{0.400pt}}
\multiput(614.00,790.59)(0.821,0.477){7}{\rule{0.740pt}{0.115pt}}
\multiput(614.00,789.17)(6.464,5.000){2}{\rule{0.370pt}{0.400pt}}
\multiput(622.00,795.59)(0.671,0.482){9}{\rule{0.633pt}{0.116pt}}
\multiput(622.00,794.17)(6.685,6.000){2}{\rule{0.317pt}{0.400pt}}
\multiput(630.00,801.59)(0.581,0.482){9}{\rule{0.567pt}{0.116pt}}
\multiput(630.00,800.17)(5.824,6.000){2}{\rule{0.283pt}{0.400pt}}
\multiput(637.00,807.59)(0.821,0.477){7}{\rule{0.740pt}{0.115pt}}
\multiput(637.00,806.17)(6.464,5.000){2}{\rule{0.370pt}{0.400pt}}
\multiput(645.00,812.59)(0.671,0.482){9}{\rule{0.633pt}{0.116pt}}
\multiput(645.00,811.17)(6.685,6.000){2}{\rule{0.317pt}{0.400pt}}
\multiput(653.00,818.59)(0.710,0.477){7}{\rule{0.660pt}{0.115pt}}
\multiput(653.00,817.17)(5.630,5.000){2}{\rule{0.330pt}{0.400pt}}
\multiput(660.00,823.59)(0.671,0.482){9}{\rule{0.633pt}{0.116pt}}
\multiput(660.00,822.17)(6.685,6.000){2}{\rule{0.317pt}{0.400pt}}
\multiput(668.00,829.59)(0.821,0.477){7}{\rule{0.740pt}{0.115pt}}
\multiput(668.00,828.17)(6.464,5.000){2}{\rule{0.370pt}{0.400pt}}
\multiput(676.00,834.59)(0.671,0.482){9}{\rule{0.633pt}{0.116pt}}
\multiput(676.00,833.17)(6.685,6.000){2}{\rule{0.317pt}{0.400pt}}
\multiput(684.00,840.59)(0.710,0.477){7}{\rule{0.660pt}{0.115pt}}
\multiput(684.00,839.17)(5.630,5.000){2}{\rule{0.330pt}{0.400pt}}
\multiput(691.00,845.59)(0.671,0.482){9}{\rule{0.633pt}{0.116pt}}
\multiput(691.00,844.17)(6.685,6.000){2}{\rule{0.317pt}{0.400pt}}
\multiput(699.00,851.59)(0.821,0.477){7}{\rule{0.740pt}{0.115pt}}
\multiput(699.00,850.17)(6.464,5.000){2}{\rule{0.370pt}{0.400pt}}
\multiput(707.00,856.59)(0.671,0.482){9}{\rule{0.633pt}{0.116pt}}
\multiput(707.00,855.17)(6.685,6.000){2}{\rule{0.317pt}{0.400pt}}
\multiput(715.00,862.59)(0.710,0.477){7}{\rule{0.660pt}{0.115pt}}
\multiput(715.00,861.17)(5.630,5.000){2}{\rule{0.330pt}{0.400pt}}
\multiput(722.00,867.59)(0.671,0.482){9}{\rule{0.633pt}{0.116pt}}
\multiput(722.00,866.17)(6.685,6.000){2}{\rule{0.317pt}{0.400pt}}
\multiput(730.00,873.60)(0.774,0.468){5}{\rule{0.700pt}{0.113pt}}
\multiput(730.00,872.17)(4.547,4.000){2}{\rule{0.350pt}{0.400pt}}
\end{picture}

\bigskip

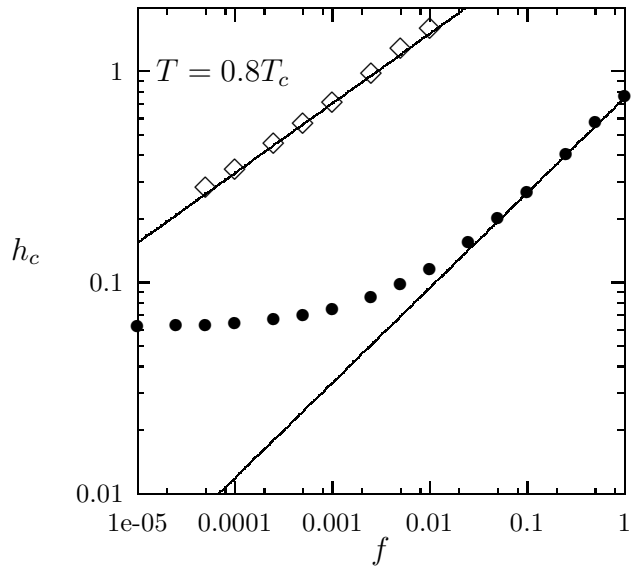
\noindent {\bf Fig.2.} Frequency ($f$) variations of 
dynamic coercive fields ($h_c$).
($\Diamond$) Monte Carlo results and ($\bullet$) meanfield results. Solid lines
are linear best fit.
\end{document}